\documentclass[mathleft,
]{an}
\usepackage{graphicx}
\usepackage{amssymb,amsmath}
\usepackage{times}
\begin{document}

\Pagespan{789}{}
\Yearpublication{2006}%
\Yearsubmission{2005}%
\Month{11}%
\Volume{999}%
\Issue{88}%

\title{Unifying neutron stars: getting to GUNS}

\author{ A.P. Igoshev\inst{1}
\and
S.B. Popov\inst{2}\fnmsep\thanks{Corresponding author:
  \email{sergepolar@gmail.com}\newline}
\and R. Turolla\inst{3,4} }
\titlerunning{Unifying neutron stars}
\authorrunning{A.P. Igoshev, S.B. Popov \& R. Turolla}
\institute{ Saint Petersburg State University, Sobolev
Astronomical Inst., Universitetski pr. 28, 198504, Saint
Petersburg, Russia \and Lomonosov Moscow State University,
Sternberg Astronomical Institute, Universitetski pr. 13, 119991
Moscow, Russia \and University of Padova, Department of Physics
and Astronomy, via Marzolo 8, Padova, Italy \and
Mullard Space Science Laboratory, University College London, Holmbury St. Mary, Dorking, Surrey, RH5 6NT, UK}
\received{30 May
2005} \accepted{11 Nov 2005} \publonline{later}


\abstract{%
The variety of the observational appearance of young isolated
neutron stars must find an explanation in the framework of some
unifying approach. Nowadays it is believed that such scenario must
include magnetic field decay, the possibility of magnetic field
emergence on a time scale $\lesssim 10^4$--$10^5$~yrs, significant
contribution of non-dipolar fields, and appropriate initial
parameter distributions. We present our results on the initial
spin period distribution, and suggest that inconsistences between
distributions derived by different methods for samples with
different average ages can uncover field decay or/and emerging
field. We describe a new method to probe the magnetic field decay
in normal pulsars. The method is a modified pulsar current
approach, where we study pulsar flow along the line of increasing
characteristic age for constant field. 
Our calculations,
performed with this method, can be fitted with an exponential
decay for ages in the range $8\times 10^4$--$3.5 \times 10^5$~yrs
with a time scale $\sim 5 \times 10^5$~yrs. We discuss several
issues related to the unifying scenario. At first, we note that
the dichotomy, among local thermally emitting neutron stars,
between normal pulsars and the Magnificent Seven remains
unexplained. Then we discuss the role of high-mass X-ray binaries
in the unification of neutron star evolution. We note, that such
systems allow to check evolutionary effects on a time scale longer
than what can be probed with normal pulsars alone. We conclude
with a brief discussion of importance of discovering old neutron
stars accreting from the interstellar medium. } \maketitle

\section{Introduction}

 Two things awe us most, initial properties of neutron stars in the
starry sky above us and magnetic field evolution within them.

Neutron stars (NSs) appear in great variety, even restricting to
isolated relatively young 
objects (age $\lesssim 1$ Myr): radio pulsars (PSRs), anomalous X-ray
pulsars (AXPs) and soft gamma-ray repeaters (SGRs), the
Magnificent Seven (M7), central compact objects in supernova
remnants (CCOs in SNRs), rotating radio transients (RRATs), etc.
(see recent reviews by \cite{harding, sandro}). These sources are
observed at all wavelengths. Their periods cover a range more than
four orders of magnitude wide, and their dipole magnetic fields
span more than six orders of magnitude. Some are observed due to
their bursting activity, some others due to their persistent
emission, which can be thermal or/and non-thermal. In addition,
transitions between different types of activity, or combinations
of different features are observed.

A question arises: {\it ``Why the good God had opened up so many
choices and made life so strange and diverse?''} (John Cheever,
``Clementina''). In other words, why NSs are observed in so
different flavors ? Can we explain all these objects in the
framework of one coherent picture without ``epicycles'' ?

 There is a hope that we are on the way towards what was called by
\cite{kaspi2010} the Grand Unification of neutron stars (GUNS
hereafter). The idea is to find a combination of initial
distributions and evolutionary laws that allows to unite all
known types of sources in one general picture, to explain all of them in one framework. 
This must also include transitions between different types of activity and
appearance of hybrid behavior (which can be called ``centaurus behavior'',
--- 
for example, PSR and magnetar at the same time, --- similar to centaurs objects
in the Solar system, which typically behave with characteristics of both
asteroids and comets).

In the first place, this approach must include non-trivial magnetic
field evolution which allows transitions between different types
of objects (or/and different types of activity). In the framework
of magnetic field decay a few first steps towards GUNS have been
made in \cite{popov2010}. After inclusion of an emerging magnetic
field,
--- a concept which became popular in last two years,--- further
advances have been made by \cite{pvg2012}. More recently, a
unified model was presented by \cite{v2013}. Still, several
phenomena lack natural explanation in the framework of GUNS.

\section{Initial spin periods}

 Initial distributions of  NS parameters are by themselves
important elements of GUNS. However, these distributions cannot be
obtained from observations directly. Typically, they are derived
using various assumptions, which in turn can be related to GUNS (for
example, the assumption of constant magnetic field conradicts GUNS).
So, comparing
distributions determined by different methods (and, probably, for
different sets of sources) we can check the assumptions made, and
obtain additional information about elements of the general
picture. In this section we are going to illustrate this issue by
discussing initial spin period distributions.

\subsection{Neutron stars in supernova remnants}

 To get an estimate of the initial spin period, $p_0$, it is necessary to
know how a NS is spinning down and to know its age. As for the age
there are several ways to have a good guess. Leaving aside
historical SN, which are few, the best way is to find a NS in a
SNR. for which it is much easier to get an age estimate. There are
tens of proposed associations NS+SNR. For several well-studied
cases initial spin periods have been derived in
\cite{migliazzo2002}.

 In \cite{pt2012a} we collected data from the literature about 30 NS+SNR
associations. For more than 20 of them it was possible to obtain
reasonable estimates of $p_0$ under the assumption of
magneto-dipole spin down with constant magnetic field (braking
index $n=3$):
\begin{equation}
p_0=p\sqrt{1-t/\tau},
\label{eq1}
\end{equation}
where $p$ is the current period, $t$ is the true age (the SNR age
in our case), and $\tau$ is the spin-down (or characteristic) age.
 Results are shown in Fig. 1.

\begin{figure}
\includegraphics[width=\hsize]{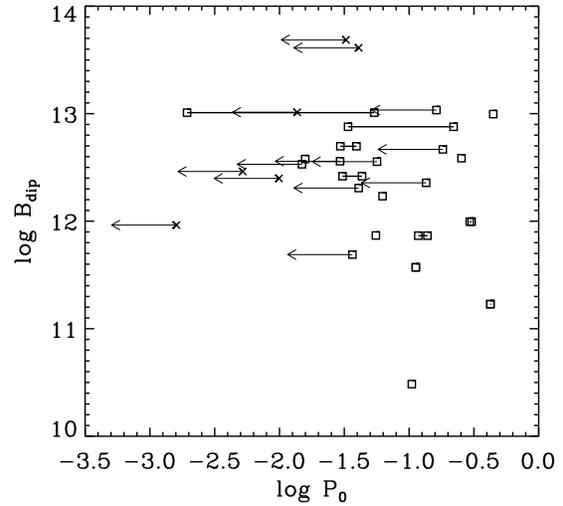}
\caption{Initial spin periods vs. magnetic field (derived from spin-down
according to the magneto-dipole formula). For some sources a range of $p_0$
is indicated. In several cases just upper limits are given. Crosses mark the
objects for which SNR ages have just upper limits compatible with spin-down
ages of PSRs. From Popov \& Turolla (2012a).}
\label{label1}
\end{figure}

The low number of objects with well-determined $p_0$ does not
allow us to produce a trustable distribution. Still, we can do the
opposite thing: to check popular analytical distributions against
our data. Such comparison demonstrates that very narrow or very
wide (for example, flat, or flat in log-scale) distributions do
not fit. On the other hand, often used gaussians with typical
values of $\langle p_0 \rangle\sim 0.1$~s and
$\sigma_\mathrm{p_0}\sim 0.1$~s fit well.

\subsection{Kinematic ages and initial spin periods}

The association with a SNR with known age is not the only
possibility to have an independent estimate of a NS age.
\cite{noutsos} used kinematic ages of NSs to derive initial spin
periods (also under the standard braking index assumption, $n=3$).
Having $>50$ kinematic age estimates these authors obtained $p_0$
for $>30$ PSRs.  Results appear to be not in full correspondence
with those by \cite{pt2012a}. The distribution of $p_0$ obtained
by \cite{noutsos} appears to be bimodal. In addition to a
gaussian-like ``standard'' part at low ($\lesssim$ few hundred
milliseconds) periods there is a ``tail'' or a second mode at
$p_0\sim 1$~s. How one can explain the difference between
distributions obtained by \cite{pt2012a} and \cite{noutsos}? Here
we focus on one possibility (see Discussion for another
possibility).

\begin{figure}
\includegraphics[width=\hsize]{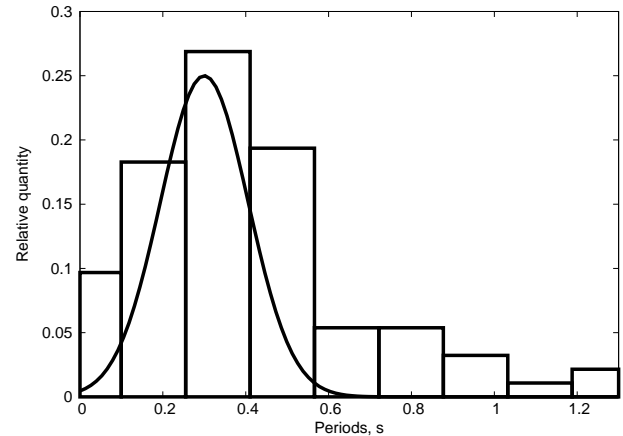}
\caption{Solid line shows the initial spin period distribution
used in our population synthesis. The histogram corresponds to the
reconstructed initial spin period distribution. A ``tail'' at
large periods appear due to magnetic field decay, which was not
included into the reconstruction assumptions. Only objects with
$10^5<t<10^7$~yrs are used for $p_0$ reconstruction. From Igoshev \& Popov
(2012).}
\label{label2}
\end{figure}

 The key point is related to the fact that the sample from \cite{pt2012a}
is nearly  two orders of magnitude younger than the sample from
\cite{noutsos}. Both reconstructions assume that there is no
effective field evolution\footnote{In the standard magneto-dipole model the
effective magnetic field is $B \sin \chi$, 
where $\chi$ is the angle between spin and magnetic dipole axis.
Often it is impossible to distinguish if the dipole field is evolving, or
the angle is changing, so we speak about the effective field evolution.}. 
For the younger sample this assumption
seems to be reasonable, since it contains no highly magnetized
sources. However, for the sample analyzed by \cite{noutsos} such a
conservative hypothesis is not so obvious: even for standard
magnetic fields $\sim 10^{12}$~--~$10^{13}$~G  field evolution can
be influential on a time scale $\sim$ several Myrs.

 If the effective 
magnetic field decayed significantly, the
 current spin-down rate is lower than in the past, and the spin-down age (for the same
true age and initial parameters) is longer than in the case of
constant field. So, eq.(\ref{eq1}) produces an overestimated
initial spin period. Appearance of such effect can be easily
demonstrated with a population synthesis calculation
\cite{ip2013}. It is necessary to specify some smooth 
initial period distribution, include magnetic
field decay in the model, and run the code to produce a population
of ``artificially observed'' radio pulsars. Then, using
eq.(\ref{eq1}) we reconstruct the initial spin period distribution
and compare it with the one used in calculation. The difference is
mainly due to the magnetic field decay.

 Results of this approach are shown in Fig. 2. The solid line gives the actual
initial spin period distribution used in the population synthesis.
The histogram corresponds to the reconstructed initial
distribution. Clearly, a ``tail'' appears in the reconstructed
distribution due to field decay, which was not accounted for in
the reconstruction.

\section{Field decay: a new approach and results}

Analysis of two initial spin period distributions reconstructed
with two different methods gives some arguments in favour of
magnetic field decay in normal radio pulsars on a time scale
$\sim$ a few million years. Can we do better using large
statistics? Yes, but we need a new method of analysis.

\begin{figure}
\includegraphics[width=\hsize]{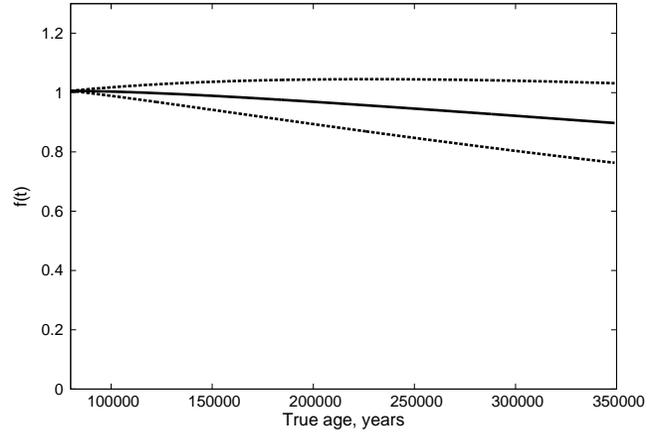}
\caption{Magnetic field decay derived from analysis of 50 samples
generated by our population synthesis code with constant magnetic
field. The solid line shows the average of all 50 samples, and the
dashed lines correspond to the variance. Some systematics (false
decay) is visible, this effect will be discussed elsewhere
(Igoshev et al., in preparation).} 
\label{label3}
\end{figure}

\begin{figure}
\includegraphics[width=\hsize]{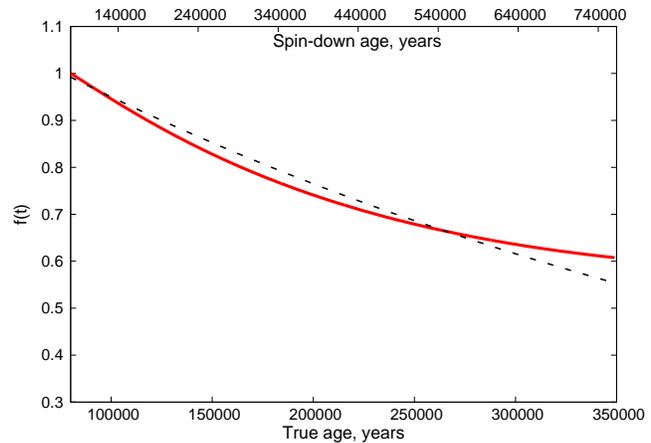}
\caption{Magnetic field decay derived from the analysis of the
sample of normal PSRs from the ATNF catalogue. The solid line
corresponds to the reconstructed decay function, and the dashed
line is an exponential fit. On the bottom horizontal axis we show
the true (statistical) age, while the spin-down age is indicated
on the top horizontal. The reconstructed law of the field
evolution is corrected for the systematic error (see the text).}
\label{label4}
\end{figure}

The original method was proposed by \cite{vn1981} and developed in
\cite{no1990}. A recent discussion can be found in \cite{vm2011}.
The basic idea is the following. Normally, the spin-down age of
PSRs, $\tau$, increases (we do  not consider young sources,
for which magnetic field emergence
can be important, see \cite{bernal2013,pt2012b} and references
therein). This happen due to spin-down and effective magnetic
field evolution. If there is no evolution of effective 
magnetic fields then the number of pulsars with spin-down age less
than some $\tau_1$ should be the same as the number of pulsars
with true age less then $\tau_1 - \overline \tau_0$, where
$\overline \tau_0$ is the average initial spin-down age for the
ensemble under study. This statement should be valid for any
$\tau_1 \gg \overline \tau_0$. The true age can be statistically
estimated, and so we have a function $t$ vs. $\tau$. When this
estimate is done, we can check if the assumption of constant field
is valid.

 We reconstruct the function $t$ vs.
$\tau$ for spin-down ages $\sim 10^5$~--~$10^6$~yrs and use it to
estimate the rate of field decay in the framework of
magneto-dipole spin-down with evolving field. To derive
$t$~--~$\tau$ from the observational data we make some
assumptions. At first, we use the range of spin-down ages in which
selection effects are not very important (this assumption is
checked by comparison of cumulative distance distributions for
sources of different ages; this assumption is mainly
related to the upper limit of the range). Then  we assume that
PSRs have some maximum spin-down age at birth (i.e., initial
positions of PSRs in the $p$-$\dot p$ diagram are confined in some
limited region), it determines the lower limit of the range of
spin-down ages used in our approach.
Finally, we assume that the birthrate of PSRs is constant.
The latter assumption allows us to introduce a ``statistical
age'', $t_\mathrm{stat}=N(\tau)/NSBR$, where $N(\tau)$ is the
number of PSRs with spin-down ages below a given value, and $NSBR$
is an internal parameter of the model which corresponds to the
birth rate of NSs used for our estimates of field decay.  We use
$t_\mathrm{stat}$ as an estimate of the true age of a PSR. So, the
subscript is dropped in the following.

The determination of $NSBR$ is related to the maximum spin-down
age at birth, $\tau_0$. We assume that for this value $t_\mathrm{stat}$
and $\tau$ are equal, so
$NSBR=N(\tau_0)/\tau_0$. This value is different from the actual
total birth rate of NSs.

After we are able to reconstruct from observations (or using data
from a synthetic model) the dependence
$t$-$\tau$, we use it to derive the function, $f(t)$, which
describes the field decay. It is assumed that the field is only
diminishing:

\begin{equation}
B(t)=B_0  f(t).
\label{eq2}
\end{equation}

In Fig.\ref{label3} we show results in which as input data we used
population synthesis calculations with constant magnetic field.
There is some systematic error, and there is some
variance due to limited statistics, however, the method
successfully reconstructs the field behaviour. The systematic
error was studied in details using population synthesis modeling
where the law of magnetic field evolution is known. We are able to
correct our results to reduce this systematics.

Then we apply our method to real data from Parkes and Swinburne
surveys. Results are given in Fig.\ref{label4}. Our calculations
demonstrate that the field is decaying for the spin-down age range
$8\times 10^4$--$10^6$~yrs (corresponding to the true age range
$8\times 10^4$--$3.5\times10^5$~yrs). The decay function can be
fitted with an exponential with time scale $4.6\times10^5$~yrs
(Igoshev et al., in preparation). The rate of decay is compatible
with the Hall time scale in normal pulsars.

\section{Discussion}

 In this section we present several GUNS-related issues, which
provide links with other types of sources, not discussed above.
Still, all of them a related to the evolution with changing magnetic field.

\subsection{``One second'' problem}


Here we present and discuss new unpublished results, related to the unified
description of the NS population in the framework of decaying magnetic
field.

 When one is developing such general approach as GUNS,
it is very important to use as many ways to compare calculations with
observations as possible. Confronting modeled data with additional observed
parameters can bring new questions, new problems, and in this subsection we
are going to discuss one.

 In (\cite{popov2010}) the authors were able to explain numbers of observed
sources of different kind (PSRs, magnetars, M7) using one framework. 
Some assumptions used in these calculations are now independently verified. 
Unique initial spin period distribution for different NSs is supported with
the data on NSs in SNRs (Sec. 2). The existence of moderate 
field decay in normal pulsars is confirmed (Sec. 3). 
On the other hand, field is
not decaying much on the time scale $1-10$~Myrs (Sec. 4.2).
However, more detailed comparison with the data shows,
that, probably, further improvements in the model are necessary.

\begin{figure}
\includegraphics[width=\hsize]{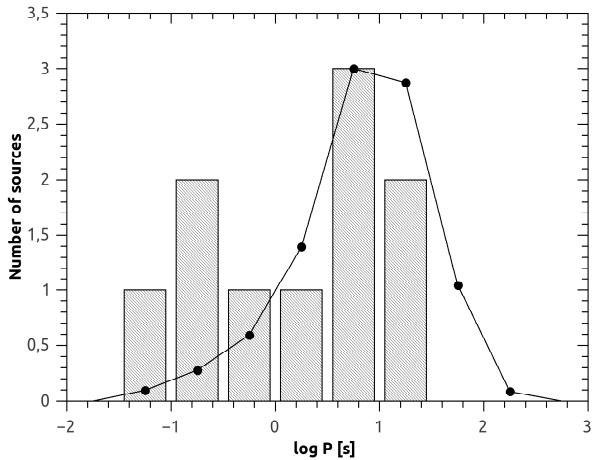}
\caption{Spin period distribution for close-by cooling NSs. The
histogram corresponds to the observed sources. Two maxima (PSRs
and M7) are clearly visible. The solid line with symbols shows the
results of calculations (Popov, in preparation).} \label{label5}
\end{figure}

Speaking about close-by cooling NSs (M7 and PSRs)
observed by ROSAT we can look at their current distribution
in the $p$-$\dot p$ diagram.
 Observed sources are divided into two well-separated groups: standard field
PSRs with $p\lesssim 0.3$~s and M7 with larger  $\dot p$
and with $p\gtrsim 3$~s. Calculations provided quite a different picture. In
Fig.\ref{label5} we present preliminary results for the data set similar to
that used in (\cite{popov2010}), and confront them with the observational
data. The synthetic distribution is smooth. Sources with $p\sim 1$~s   are
predicted, but are not observed.

 Several explanations can be proposed. The first is obvious: we have very
low observational statistics. Still, the fact that
sources in different peaks belong to different subpopulation of
NSs is against it. The second explanation is related to unmodeled
(and unknown) selection effects. Indeed, the underlying
distribution can be smooth (as predicted by the model), but in
observations we see two separated groups. However, preliminary
analysis of possible effects does not allow us to fit the data
(Popov, in prep.). Finally, it is possible that the model needs
modifications (for example, cooling of NSs with low and standard magnetic
fields can be finetuned to make contribution of such sources larger). 
At the moment, we think that this option is the
most probable, and new calculations are in progress.

Joint description of the magneto-rotational and cooling evolution of NSs
of all types in one population synthesis model would be the final step for
GUNS. But it seems that several important issues are not clear, yet.

\subsection{High-mass X-ray binaries and NS evolution}

GUNS cannot be limited to isolated NSs alone. Binaries inevitably
have to be included. This can be tackled from two sides. First,
some types of binaries  are excellent test beds for NS evolution.
Second, if we have at least some general ideas about GUNS, then we
can use them to explain properties of peculiar sources in
binaries. In this subsection we give examples for both approaches.


Magnetic field evolution is normally tested using data on PSRs and
magnetars. This means that for highly magnetized NSs we cannot
confront predictions vs. observations on a time scale larger than
few tens or hundreds thousand years.  High-mass X-ray binaries
(HMXBs) give an opportunity to solve this problem. NSs in these
systems have ages $\gtrsim$1--10 Myrs. Since accretion normally is
due to stellar wind, the accretion rate is not very high and
duration of the accretion stage is not very long, 
so the field is not much influenced by it. 
Determination of magnetic fields is possible in rare cases
directly via cyclotron line observation. But mostly fields can be
estimated using known spin periods and their variations.

 Several methods of magnetic field estimation in X-ray pulsars
were applied by \cite{chashkina}.  Among them the authors used a
new model of quasi-spherical accretion by \cite{shakura}.
Application of this model allowed to demonstrate that magnetic
field distribution in HMXBs is compatible with predictions of the
scenario by the Alicante group (see, for example,
\cite{aguilera}). 
For standard-value fields the distribution
of NSs in HMXBs is similar to that of PSRs, 
i.e. no additional significant
decay happens during lifetime of HMXBs. 
 We think that HMXBs can be fruitfully used for further comparison of
the GUNS prediction with observational data.


Now, we briefly illustrate how standard ingredients of GUNS can be
used to explain properties of peculiar sources in binary systems.
 We do this by looking at SXP1062 --- a recently discovered X-ray binary in the
SMC.

 The unique feature of  SXP1062 is its association with a SNR. This
provides an estimate for the age of the NS in this system, $\sim
(2$--$4) \times 10^4$~yrs (\cite{haberl,brunet}). If we assume
that the source is close to spin equilibrium (new observations
support it, \cite{sturm}), then the present day field is
$\sim10^{13}$~G. With such short age it is difficult to come to
the stage of accretion and spin-down the NS to 1062~s period. 
There are two possibilities. The first, proposed by
\cite{haberl}, is related to long initial spin period: $p_0\sim
1$s. The second, proposed by \cite{pt2012c}, is related to the
magnetic field decay.  If the latter possibility is realized, then
SXP1062 is an evolved magnetar in a HMXB system --- the first
example of such a source.

\subsection{Buried and resurrected}

  A scenario for GUNS includes the possibility that magnetic field can be
initially buried by intensive fall-back accretion, and then the
field emerges on a timescale $\lesssim 10^4$--$10^5$~yrs
(\cite{muslimov,ho}). In this subsection we briefly discuss
several cases in which this process can be important.


 The bestiary of NSs is continuously enriched with new monsters.
 A NS in the SNR Kes 79 was proposed to be a buried magnetar (\cite{lai}).
If this is the case, we have to find a place for this object in the GUNS.
Moreover, similar sources can give a clue to the formation mechanisms of
magnetars (\cite{popov2013}).

 The spin period of this source is 0.105~s. The period derivative is small, 
so the present inferred dipole field is low. However, large pulse
fraction points to large crustal field (\cite{lai}). If the field
was rapidly buried during the first minutes or hours after the NS
formation (as the standard scenario predicts), then the present
day spin period is very close to the initial one. The value
0.105~s is in contradiction with the standard mechanism of field
generation in magnetars (\cite{dt1992}). Two possibilities can be
discussed. Either, in Kes79 we have a true magnetar, and so the
dynamo mechanism is not valid. Or, the object in Kes 79 is somehow
different from normal magnetars (maybe, belonging to low-field
magnetars, see \cite{te2013}), the dipole magnetic field of which
are not too large at birth). Discovery of a similar object, but
with a millisecond period, would be a proof for the standard
Duncan-Thompson scenario.


CCOs with low-fields (the so-called anti-magnetars) are believed
to be objects with buried magnetic fields. Do we have other
evidence in favour of buried and emerging field? In our opinion,
two observations can be made to support this picture
(\cite{pt2012b}).

The first is related to close-by cooling NSs. As it was noted
already, there are two sub-populations among these sources: PSRs
and M7. Detailed modeling also shows that there is no necessity to
add more sources of different nature. However, in SNRs we see that
a significant fraction of sources belong to CCOs. If so, we expect
to see matured CCOs around us as thermal X-ray sources. Their
absence provides an indirect argument in favour of the hypothesis
that such objects ``disappear'' on a time scale $\lesssim
10^5$~yrs due to field emergence. I.e., probably there are matured
CCOs in the solar vicinity observed as soft sources, but we do not
recognize them.

For the second we have return to HMXBs.
Again, if anti-magnetars form a significant fraction of young NSs, then we
expect to find them in HMXBs, unless something happens. 
There are no confirmed NSs with low magnetic fields in HMXBs. 
This also can be considered as an
indirect argument in favour of emerging magnetic field.

Finally, the difference between initial spin period distributions
derived by \cite{pt2012a} and \cite{noutsos} can also be explained
by emerging magnetic field. NSs from the ``tail'' in
\cite{noutsos} could be absent in the younger sample by
\cite{pt2012a}.
Simply, sources visible in the older population in the ``tail''
could be ``hidden'' by fall-back in their early ages.

\subsection{Alignment}


Often when we discuss magnetic fields of NSs (especially, in the
context of magneto-rotational evolution) we mean effective field,
which includes also the angle between spin and dipole axis. 
Evolution of this angle towards the position of smaller energy
losses can mimic magnetic field decay. Probably, this is one of
the most elusive (luckily, also one of the least important)
ingredient of GUNS.

Potentially, HMXBs can be used to measure the angle and to put limit on its
evolution (as they can
be used to test field evolution on the time scale
$\gtrsim 1$~--~$10$~Myrs). A preliminary analysis (\cite{karino})
shows that  angles are not close to 0 or 90 degrees in
sources with average ages about few Myrs, which, in our
opinion, excludes significant evolution of the angle on shorter
(pulsar life time) time scales. Additionally, in the early version
of a compilative Be/X-ray binaries catalogue (\cite{pr2004}) we
look for correlation of pulse fraction with other parameters of
the sources. Nothing was found, and potentially this argues
against significant angle evolution towards one of limiting
positions. Future theoretical studies 
which include detailed models for pulse shape are necessary.

\subsection{Isolated accretors}

 A major step in understanding of NSs will be taken when really old isolated
objects are discovered. Probably, the unique possibility to do it is to find
isolated accreting NSs. This will allow us to test models
on a time scale $\sim$ few Gyrs.

 Accreting isolated NSs have been discussed since early 70s. Some hopes to
detect them were related to ROSAT (see a review in \cite{treves}).
Then it was shown that in a standard (at that moment) evolutionary
scenario NSs mostly do not reach the stage of accretion from the
interstellar medium (\cite{census}). However, modern scenario makes
predictions slightly more optimistic.

\cite{bp2010} used analytical approximations to the evolutionary
scenario from (\cite{aguilera}) to model NS evolution on a 
long time scale. It was shown that NSs with initially large
magnetic fields are
able to reach the stage of accretion. Funny thing about it is the
following. Despite optimistic predictions in early 90s, no
accreting isolated NSs were discovered. Instead, unpredicted
cooling NSs (M7) were found.  But
exactly these sources in future may become isolated accretors, and
if there are accreting isolated NSs around us most probably they
were M7-like when they were young!

Discovery of old (accreting) isolated NSs is very important for GUNS.
This is a task for future
surveys in soft X-rays, for example for eROSITA onboard Spectrum-X-Gamma,
which can also contribute to young NSs studies.

\section{Conclusions}

During the last several years the zoo of young isolated NSs
started to look not so unexplainably motley. Some evolutionary
links between different types of sources are established, and more
are coming with the help of the concept of emerging magnetic
field. Different approaches to model and interpret data can help
to probe various assumptions, as it probably happened with the
initial spin period distribution. Diverse methods to confront
model predictions with observations are necessary, because
evolutionary scenarios now have too many ingredients and free
parameters. Inclusion of HMXBs in these considerations can be
fruitful in many respects, especially probing magnetic field and
spin-dipole angle evolution. Discoveries of isolated NSs beyond
magnetar, radio pulsar, and residual thermal emission stages is
very much welcomed.

\acknowledgements
We thank the organizers of the conference for hospitality and participants
for discussions. A.I. thanks the SPbU grant 6.38.73.2011,
S.P. thanks the RFBR (grant 12-02-00186).



\end{document}